\documentclass[aps, nofootinbib]{revtex4}
\usepackage{filecontents}
\usepackage[T1]{fontenc}
\usepackage{mathrsfs}
\usepackage{amsmath}
\usepackage{amssymb}
\begin{document}
%\title{\textbf{Unlocking the duality between exotic oscillators and dissipative system: Shadow of Newton-Hooke symmetry}}

\title{Symmetry Duality: Exploring Exotic Oscillators And Dissipative Dynamics Through The Glass Of Newton-Hooke}	
\author{Sayan Kumar Pal $^{1}$}\email{pal.sayan566@gmail.com}
 \author{Partha Nandi $^{2}$}\email{pnandi@sun.ac.za}
 \affiliation{$^{1}$Physics and Applied Mathematics Unit, Indian Statistical Institute, Kolkata, 203 B.T. Road, West Bengal 700108, India.}
\affiliation{$^{2}$Department of Physics,
University of Stellenbosch, Stellenbosch-7600, South Africa.}
%\author{Sayan Kumar Pal\footnote{pal.sayan566@gmail.com}, Partha Nandi\footnote{pnandi@sun.ac.za}\\
%^{a} Physics and Applied Mathematics Unit, Indian Statistical Institute, Kolkata,\\
%203 B.T. Road, West Bengal 700108, India\\
%^{b} Department of Physics,\\
%University of Stellenbosch, Stellenbosch-7600, South Africa}

%\title{Unlocking the duality between exotic oscillators and dissipative system: Shadow of Newton-Hooke symmetry}\\

 % % % % % % % % % % % % % % % % % % % % % % % % % % % % % % % % % % % % % % % % % % % % % % % % % % % % % % % % % % % % % % % % % % % % % % % % % % % % % % % % % % % % % % % % % % % % % % % % % % % % % % % % % % % % % % % % % % % % % % % % % % % % % 
\begin{abstract}
Motivated by the symmetry in the non-relativistic limit of anti-de Sitter geometry, we employ planar dynamical models featuring exotic (deformed) harmonic oscillators, presented through direct and indirect Lagrangian representations. The latter introduces Bateman dissipative oscillator system. Analyzing these dynamic systems with a first-order Lagrangian scheme, our phase-space-based approach utilizes the moment map components to reveal the underlying symmetry algebra. This obtained algebra, interpreted as an extended version of Newton-Hooke (NH) cosmological symmetry algebras, has the potential to cast an augmented non-relativistic shadow over the expanding universe, offering an insightful perspective on extended NH spacetime in 2+1 dimensions through our dynamical realizations.
\end{abstract}

\maketitle

%\documentclass[12pt,a4paper]{article}
%\usepackage[latin2]{inputenc}

%\usepackage{slashed}

 % % % % % % % % % % % % % % % % % % % % % % % % % % % % % % % % % % % % % % % % % % % % % % % % % % % % % % % % % % % % % % % % % % % % % % % % % % % % % % % % % % % % % % % % % % % % % % % % % % % % % % % % % % % % % % % % % % % % % % % % % % % % % 

\section{Introduction}
Studies in nonrelativistic spacetimes with a special emphasis on their associated symmetries have particularly gained importance and popularity due to some exciting discoveries in recent times. These studies have provided a foundation for studies of transport phenomena in nonrelativistic fluids \cite{prabhu}, Chern-Simons gravity theories \cite{obers}, phonons moving in a perfect nonrelativistic fluid \cite{visser}, nonrelativistic Nambu-Goldstone modes \cite{kobayashi}, Newton-Cartan gravity and supergravity theories \cite{rbnew, bergshoeff}, Dirac fermions in curved space-times \cite{annalenderphysik}, etc. All these studies are broadly directed towards one of the two major themes: the AdS-CFT correspondence \cite{sachdev} and the analogue gravity theories \cite{barcelo, nat}. At this point, we would want to clarify to the readers that by space-times, we may mean here even a degenerate metric between space and time, which is typically the case in nonrelativistic scenarios. As a matter of fact, nonrelativistic spacetimes of relevance in the present context such as Newton-Hooke spacetimes or the similar Galilei-Maxwell spacetime serve as examples of non-metric but curved spacetimes as their canonical connections have non-vanishing Riemann curvature tensors, in contrast to the Galilean space-time. Associated with these nonrelativistic spacetimes, one has the corresponding symmetry groups which have become a major thrust of research in recent times due to their cosmological and statistical mechanical applications \cite{ellis, duval2017, dtson, martelli}, of which the Newton-Hooke group in (2+1) dimension ($\mathcal{NH}_3$) becomes important in our present context. In fact, the Newton-Hooke (NH) group in n+1 dimensions is the symmetry group of the n-dimensional harmonic oscillator (H.O.) and can be obtained by a group contraction of the anti-de Sitter group in n+1 dimensions, $AdS_n$ \cite{zhang1, zhang2, gibbons1, NHspacetime1}. Newton-Hooke spacetimes have been interpreted as flat and torsionless Newton-Cartan spacetime with a quadratic Newtonian potential and have also been described as a non-relativistic FLRW geometry with flat spatial slices \cite{grosvenor}. They are therefore considered as cosmological nonrelativistic groups, and as such may be regarded as dynamical groups of (free and nonrelativistic) Galilean systems, where the frequency of the system plays the role of a Hubble's constant in an accelerated expanding/contracting universe \cite{aldrovandi,gibbons}. Just recently in \cite{gibbonsnewest}, using intuitive arguments from Newton-Hooke theory and subsequently invoking numerical solutions of Einstein's equations, the existence of static black hole binaries has been reported in de Sitter spacetime thereby providing the first four-dimensional example of nonuniqueness in general relativity without matter. A classification of systems in (2+1)D, both classical and quantum, having Newton Hooke symmetry has been presented in \cite{olmo}, which has also discussed the exotic central extensions of the group in (2+1) dimensional context. Here we aim to provide a dynamical realization of the extended Newton-Hooke group $\overline{\mathcal{NH}_3}$, where we borrow the notation from \cite{olmo} mentioned just above. This extended Newton-Hooke group $\overline{\mathcal{NH}_3}$ will be a nine-dimensional group containing three central extensions. In addition to that, the extended $\overline{\mathcal{NH}_3}$ resembles partly the reduced Galilei-Maxwell group corresponding to the magnetic Landau group limit \cite{negro06,negro2002}. In \cite{newNH4}, a study of (1+1)D Newton-Hooke group for the simple and the damped H.O. has been carried out within the framework of the orbit method, while here we present a study corresponding to the more physical and interesting (2+1)D case using moment maps \cite{simalg, cariglia} which, roughly speaking, are generalizations of the notion of Noether charges in a symplectic manifold.

Besides, there have been studies that indicate non-trivial geometries or space-times to be associated with the noncommutativity of dynamical operators at the quantum level or symplectic deformations of phase space degrees of freedom at the classical level, see for example \cite{shahn,glik101,glik102,cortes,pjhep}. Alternatively, the presence of a Berry curvature in momentum space deforms the phase-space bracket relations \cite{xiao}. This motivates us to present here a symmetry study of deformed systems by taking the exotic oscillators \cite{say2,stichel,sir2} as our basic system, which is basically a 2D H.O. embedded in a deformed symplectic space where position coordinates satisfy a non-zero Poisson structure. The deformed oscillators are thought to represent a low-energy nonrelativistic effective system that is believed to be originally derived from physics near Planck scales. More significantly, from the symplectic viewpoint, deformations in the symplectic structure naturally arise inside solids when electron dynamics is described in terms of Wannier coordinates and Bloch wavefunctions \cite{xiao}. It is worth mentioning at this point that there exists a mapping between exotic oscillators and the planar motion of an electron in a normal magnetic field together with a harmonic trapping \cite{say2}. The related problems of the planar Landau quantum systems have attracted much interest over the last few decades as they display integer and fractional quantum hall effects, which is debatably the first and the simplest manifestation of the topological phase of quantum matter. In this light of view, symmetry studies of these kinds of fundamental systems become very important which shed deep clues into many intriguing features and therefore have been undertaken in \cite{negro2002, Gomis, geracieson}, etc. Indeed, as we now well know, the Galilean group in 2+1 dimensions admits quite peculiarly two central extensions - mass and anyonic spin \cite{levy}. These 2D anyonic systems, arising in the effective description of the fractional quantum Hall effect, have also been given a symplectic phase-space description in terms of noncommutativity in the position variables and such a description can readily give rise to the Laughlin wavefunctions \cite{ady}. Without much ado, we return to our prime focus and remind that the Newton-Hooke group is distinctively characterized by noncommutativity between time and spatial translation generators, unlike in the Galilean algebra. Thus, analogously to the (anti) de Sitter cases, energy is not invariant under spatial translations. Now here in the present investigation, we will show that the boosts don't commute which can be ascribed to the anyonic spin \cite{jackiwnair,duval}. Moreover, the obtained symmetry algebra strikingly contains spatial translation generators that also no longer commute, a fact reminiscent of the anti-de Sitter or the similar Galilei-Maxwell algebra, and suggests the existence of a non-zero curvature. \\

Now, dissipation is a ubiquitous aspect of real-life physical systems. Even so, dissipation is believed to play a crucial role in the physics of quantum systems around Planck scales \cite{hoo,hooftar}. If general relativity is true at all scales, black holes that absorb information inevitably arise in the description of physics at very small scales viz at the Planck scale. Thus speculations have always been there on the role of dissipation at such high energy scales \cite{hooftbook, visco, mairi, ameli}. Furthermore, in \cite{skp1}, an interplay between dissipation and deformation has been observed through the path-integral scheme. Therefore, it becomes interesting to have a comparative study from the symmetry perspective between these two systems, which has not been undertaken previously. Thus, the above considerations make our purpose two-fold in the present context. In this paper, we show at the classical level the emergent damping mechanism in an otherwise conservative system of 2D harmonic oscillators (H.O.) in the so-called indirect representation. This observation is then strengthened by studying the symmetry algebra of the relevant Bateman oscillators (a canonical model for treating dissipative mechanics) \cite{bate} and the exotic (deformed) oscillators, which opens up a fairly new door for investigating such systems. We find the symmetry algebra identical in both systems.\\
The paper is organized in the following manner: In section II, we first study the symmetry group of the exotic oscillators using moment map components. In section III, we then consider a system of 2D harmonic oscillators in indirect representation embedded in a deformed position space and show the appearance of damping and anti-damping in the two modes. Later, we analyze the symmetry of the resulting damped and anti-damped system. In section IV, we discuss how the standard extended $\overline{\mathcal{NH}_3}$ symmetry algebra can be obtained from the $AdS_3$ algebra through a suitable group contraction. Finally, we summarise our results and conclude in section V. We have also provided an appendix to recapitulate the symmetry group of the 1D H.O., which is the firstly extended Galilei algebra with the four parameters: Boost (K), momentum (T), Hamiltonian (H), and the mass (m) as the central extension. A contrast of the Galilean algebra with the Newton-Hooke algebra has been included in appendix A, while appendix B contains the Jacobi identities in terms of the redefined generators of $\overline{\mathcal{NH}_3}$ showing the algebraic consistency of the obtained symmetry algebra. 

\section{H.O. in 2D deformed space: The exotic oscillator}
We start by considering a 2D H.O. system but here we intend to introduce spatial deformation (non-commutativity) in the system along the lines of \cite{deriglazov}. For this, we work in the first-order Lagrangian formalism. 
The Lagrangian is written as:
\begin{equation}{\label{lagrangian1}}
L=p_i\dot{x}_i -\frac{p_i^2}{2m}- \frac{1}{2}m\omega^2 x_i^2 +\frac{\theta}{2}\epsilon_{ij}p_i \dot{p}_j
\end{equation}
where we treat $(x,p)$ as configuration space variables. For the sake of clarity, we provide here the constraint analysis of the above system to explicitly display the deformation of position space of the system of 2D H.O. To that end, computing the canonical momenta, we get-
$$\pi_i^{(x)} = \frac{\partial L}{\partial \dot{x}_i} =p_i$$ and,
$$\pi_i^{(p)} = \frac{\partial L}{\partial \dot{p}_i} =-\frac{\theta}{2} \epsilon_{ij}p_j$$
These yield the primary constraints and it is a 2nd class constrained system. The set of constraints identified are,
$$ \phi_i^1 =\pi_i^{(x)}-p_i \approx 0$$ and,
$$ \phi_i^2 =\pi_i^{(p)}+\frac{\theta}{2} \epsilon_{ij}p_j \approx 0$$
The constraint matrix is
\begin{equation}
C_{ij}^{\alpha \beta}=\begin{pmatrix}
0 & -\delta_{ij}\\
\delta_{ij} & \theta\epsilon_{ij} \
\end{pmatrix}
\end{equation}
We construct the Dirac bracket defined by the following equation:
 \begin{equation}
 \{M_i,N_j\}_{DB}=\{M_i,N_j\}_{PB}-\{M_i,\phi_\alpha^k \}_{PB} (C^{-1})^{\alpha\beta}_{kl} \{\phi_\beta^l,N_j\}_{PB}\label{o}
 \end{equation}
 The Dirac brackets are computed to be,
\begin{equation}
    \{x_i,x_j\}_{DB}=\theta \epsilon_{ij}~~;~~ \{x_i,p_j\}_{DB}=\delta_{ij}~;~\{p_i,p_j\}_{DB}=0
    \end{equation}
    Next, computing the Euler-Lagrange equations of motion from (\ref{lagrangian1}), we get:
\begin{equation}\label{eqn1}
\ddot{x}_i-m\omega^2\theta \epsilon_{ij}\dot{x}_j + \omega^2 x_i =0
\end{equation}
Thus, the system in (\ref{lagrangian1}) can be interpreted as a 2-dim S.H.O. in NC position space with the standard Hamiltonian-
\begin{equation}
H=\frac{p_i^2}{2m}+ \frac{1}{2}m\omega^2 x_i^2 
\end{equation}
but now with the following non-trivial symplectic structure:
\begin{equation}
\{x_i,x_j\}_{DB}=\theta \epsilon_{ij}
\end{equation}
The trajectories of the system (\ref{lagrangian1}) or equivalently (\ref{eqn1}) are given by,
\begin{equation}
x_i= \cos{\beta t}(A_i \cos{\Omega t}+B_i\sin{\Omega t})-\sin{\beta t} \epsilon_{ij}(A_j \cos{\Omega t}+B_j\sin{\Omega t}) \nonumber
\end{equation}
where, $\beta=-\frac{m\omega^2\theta}{2}$ ~,~~ and $\Omega=\omega\sqrt{1+\frac{m^2\omega^2\theta^2}{4}}=\sqrt{\omega^2+\beta^2}$.\\
Thus, the space of motion of the particular system $\mathcal{R}^4$ is spanned by the parameters $A_i, B_i$.
After a straightforward lengthy computation, the moment map components \cite{simalg, cariglia} are found to be given by,
\begin{eqnarray}
K_i =& \bigg(m \cos{\beta t} cos {\Omega t} - \frac{m\beta}{\Omega} sin{\beta t} sin {\Omega t}\bigg) x_i - \frac{1}{\Omega} \cos{\beta t} sin {\Omega t} ~p_i \nonumber \\
&+ \bigg(m \sin{\beta t} cos {\Omega t} + \frac{m \beta}{\Omega} cos{\beta t} sin {\Omega t}\bigg) \epsilon_{ij}x_j - \frac{1}{\Omega} \sin{\beta t} sin {\Omega t} ~\epsilon_{ij}p_j 
\end{eqnarray}
and,
\begin{eqnarray}
T_i =& \bigg(m\beta \sin{\beta t} cos {\Omega t} + m \Omega cos{\beta t} sin {\Omega t}\bigg) x_i + \cos{\beta t} cos {\Omega t} ~p_i  \nonumber \\
&- \bigg(m\beta \cos{\beta t} cos {\Omega t} - m \Omega sin{\beta t} sin {\Omega t}\bigg) \epsilon_{ij} x_j +  \sin{\beta t} cos {\Omega t} ~\epsilon_{ij}p_j 
\end{eqnarray}
where K's are the NH boost generators and T's are the NH spatial translation generators. Finally, we compute the Dirac brackets among the symmetry generators to arrive at the following symmetry algebra :
\begin{equation}\label{boostalg1}
\{K_i,K_j\}=\theta m^2 \epsilon_{ij} 
\end{equation}
\begin{equation}\label{transalg1}
\{T_i,T_j\}=\theta m^2 \Omega^2 \epsilon_{ij}
\end{equation}
\begin{equation} \label{newalgebra}
\{K_i,T_j\}=m(1+\frac{ m^2 \omega^2 \theta^2}{2})\delta_{ij}
\end{equation}
\begin{equation}\label{NCPoisson}
\{H,K_i\}=-T_i + \beta \epsilon_{ij}K_j~ ~;~\{H,T_i\}=\Omega^2 K_i + \beta \epsilon_{ij}T_j
\end{equation}
As all the above generators are symmetry generators $G:=\{K_i, T_i, H\}$, one may check that $\frac{dG}{dt}=0$ holds indeed. 
Note that the spatial translations do not commute here unlike in the case of 2-D harmonic oscillators and boosts also don't commute. Particularly, the noncommutativity of the boost generators is solely due to the deformation in the symplectic structure w.r.t the position variables and also arises in anyonic systems \cite{duval,stichel}. It is worthwhile to mention here that the above-obtained symmetry algebra is reminiscent of the Newton-Hooke algebra $\mathcal{NH}_3$ which we will take up in section IV. As a passing remark, we mention here that there obviously exists another conserved generator $J$ in the system, the generator of rotations in the plane which we have not mentioned in (\ref{NCPoisson}) since the commutations between $J$ and the other generators remain unchanged, which can be verified easily. Having provided an explicit realization of the form of the symmetry generators together with the symmetry algebra for the exotic oscillators, let us now dive into the second part of the paper which presents a novel analysis of 2D H.O. in indirect representation in a deformed position space and a parallel study of its associated symmetries.

\section{2D H.O. in indirect representation in symplectic deformed space: Emergence of Bateman's model of dissipation}
The damped harmonic oscillator (dho) is characterized by the following equation:
\begin{equation}\label{dampedosci}
\ddot{Y} + \gamma\dot{Y} + \omega^2Y = 0
\end{equation}
which represents a non-conservative system as the Lagrangian, or so the Hamiltonian is explicitly time-dependent. Here $\gamma>0$ and $\omega$ are independent of time representing the damping parameter per unit mass and the frequency respectively. Non-conservative systems are notorious to work with but fortunately, H. Bateman gave a canonical formulation to handle dissipative systems \cite{bate} in which one doubles the system by also considering additionally the time-reversed image of equation (\ref{dampedosci}) where the action functional for the Bateman's doubling scheme is given by:
\begin{equation}
S=\int dt L~~~;~~L = m\dot{Y_1}\dot{Y_2} + m\frac{\gamma}{2}\left(Y_1\dot{Y}_2 - \dot{Y}_1Y_2\right)
       - m\omega^2Y_1Y_2 \label{n}
\end{equation}
This form (\ref{n}) has been referred to as the indirect representation in modern literature \cite{Morse, Sant, Vit2}. The whole system is closed with damped (source) and anti-damped (sink) harmonic oscillators along $Y_1$ and $Y_2$ axes on the 2D real plane. A thorough study of the quantum mechanics of the d.h.o in the above-mentioned framework of the Bateman dual system (\ref{n}) was carried out in \cite{jizba}. The analysis of Bateman is very similar to the notion of Thermofield double (TFD) used widely in non-equilibrium thermodynamics \cite{Vit2}. Now, the case $\gamma=0$ leads us to a system of 2D H.O. in the indirect representation and becomes interesting in the present sequel. Below, we put forward a first-order Lagrangian for the two-dimensional H.O. in the indirect representation that is embedded in a symplectic-deformed position space,
 \begin{equation}\label{batemanL2}
L = \pi_1\dot{Y}_1 + \pi_2\dot{Y}_2-\frac{1}{m}\pi_1 \pi_2 - m\omega^2 Y_1 Y_2 + \frac{\theta}{2}\left(\pi_1\dot{\pi}_2- \dot{\pi}_1 \pi_2\right)
\end{equation}
where the last term is a non-relativistic Chern-Simons like term in the momentum (here $\pi_i$'s are the canonical momenta) variables and is specifically responsible for the deformation of the associated configuration space. This Lagrangian (\ref{batemanL2}) also gives rise to a second-class constrained system and the analysis follows similarly to the previous section which we therefore omit here. It is seen that the coordinates of the system described by (\ref{batemanL2}) satisfy $\{ Y_i,Y_j\}=\theta\epsilon_{ij}$, whereas the equations of motion obtained from the above Lagrangian are:
 \begin{equation}\label{bateman1}
   \ddot{Y_1} - \lambda\dot{Y_1} + \omega^2 Y_1 = 0~ ;~~~~~~\lambda=m\omega^2\theta 
 \end{equation}
 and,
 \begin{equation}\label{bateman2}
  \ddot{Y_2} + \lambda\dot{Y_2} + \omega^2 Y_2 = 0
  \end{equation} 
which interestingly represent anti-damping (if $\theta$ is chosen positive) and damping respectively. These two equations together constitute the Bateman doubled system\footnote{These pair of equations are sometimes alternatively described \cite{Vit2,mairi} in terms of light-cone coordinates $y_i=(Y_1 \pm Y_2)/\sqrt{2}$ yielding,
\begin{equation}\label{batemandir}
   \ddot{y_1} - \lambda\dot{y_2} + \omega^2 y_1 = 0~ ,~~~\ddot{y_2} - \lambda\dot{y_1} + \omega^2 y_2 = 0
 \end{equation}}
  to handle dissipative dynamics just like in the standard equation (\ref{dampedosci}) and its time-reversed image, where $\lambda=m\omega^2\theta$ plays the role of the damping parameter $\gamma$. We now look forward to extending our present analysis of the Bateman system, obtained from an appropriate symplectic deformation as discussed above, to find the symmetry properties and to further discern the similarities or dissimilarities with the exotic oscillator of the previous section.
  \subsection{Symmetry algebra for the Bateman's dual model of dissipation}
The associated trajectories for the Bateman pair (\ref{bateman1}) and (\ref{bateman2}) result in:
 \begin{equation}
Y_i= e^{\pm \frac{\lambda t}{2}} \left(A_i \cos{\tilde{\omega} t} + B_i \sin{\tilde{\omega} t} \right) ~;~~\text{$+$ for $Y_1$ and $-$ for $Y_2$} \nonumber
 \end{equation}
 %\begin{equation}
 %Y_2= e^{-\frac{\lambda t}{2}} \left(A_2 \cos{\tilde{\omega} t} + B_2 \sin{\tilde{\omega} t} \right) \nonumber
%  \end{equation}
  where $\tilde{\omega}^2= \omega^2 - \frac{\lambda^2}{4}$.~
   The components of the moment map are again computed for the present system and given to be,
\begin{equation}
K_1= e^{-\frac{\lambda t}{2}} \left(m\cos{\tilde{\omega} t}- \frac{m\lambda}{2\tilde{\omega}}\sin{\tilde{\omega} t}\right)Y_1 - e^{-\frac{\lambda t}{2}}~ \frac{\sin{\tilde{\omega} t}}{\tilde{\omega}} \pi_2 \nonumber
\end{equation}
\begin{equation}
K_2= e^{\frac{\lambda t}{2}} \left(m\cos{\tilde{\omega} t}+ \frac{m\lambda}{2\tilde{\omega}}\sin{\tilde{\omega} t}\right)Y_2 - e^{\frac{\lambda t}{2}}~ \frac{\sin{\tilde{\omega} t}}{\tilde{\omega}} \pi_1 \nonumber
\end{equation}
and,
\begin{equation}
T_1= e^{-\frac{\lambda t}{2}} \left(m \tilde{\omega}\sin{\tilde{\omega} t}+ \frac{m\lambda}{2}\cos{\tilde{\omega} t}\right)Y_1 + e^{-\frac{\lambda t}{2}}~ \cos{\tilde{\omega} t}~ \pi_2 \nonumber
\end{equation}
\begin{equation}
T_2= e^{\frac{\lambda t}{2}} \left(m\tilde{\omega}\sin{\tilde{\omega} t}- \frac{m\lambda}{2}\cos{\tilde{\omega} t}\right)Y_2 + e^{\frac{\lambda t}{2}}~ \cos{\tilde{\omega}t} ~\pi_1 \nonumber
\end{equation}
Thus, the symmetry algebra between the symmetry generators for the Bateman pair turns out as follows,
\begin{equation}\label{boostalg}
\{K_i,K_j\}=\theta m^2 \epsilon_{ij}
\end{equation}
\begin{equation}\label{transalg}
\{T_i,T_j\}=\theta m^2 \tilde{\omega}^2 \epsilon_{ij} 
\end{equation}
\begin{equation} \label{newalgebra2}
\{K_i,T_j\}=m\bigg(1-\frac{ m^2 \omega^2 \theta^2}{2}\bigg)(\delta_{i1}\delta_{j2}+\delta_{j1}\delta_{i2})
\end{equation}
\begin{equation}\label{NCPoisson2}
\{H,K_i\}=-T_i + (-1)^i \frac{\lambda}{2} K_i~ ~;~\{H,T_i\}=\tilde{\omega}^2 K_i + (-1)^i \frac{\lambda}{2} T_i
\end{equation}
Appealingly in this case too, the symmetry algebra is very similar to the algebra of symmetries of the exotic oscillator system and is another extended $\overline{\mathcal{NH}}_3$ algebra. However, at this point, one might notice slight differences in the last terms of the r.h.s of (\ref{NCPoisson2}) with (\ref{NCPoisson}). These differences stem from the indirect representation of the Bateman oscillator system in terms of the standard Bateman damped and anti-damped variables ($Y_2,Y_1$) and can be eliminated by working with the light-cone coordinates $y_1,y_2$ (see equation 19 in footnote previously, also see \cite{mairi, Vit2}). This is also the reason for the difference in the algebra between boost generator and translation where we can notice that it is $\{K_1, T_2\} \neq 0$ in (\ref{newalgebra2}) rather than $\{K_1, T_1\} \neq 0$ typical for most dynamical systems. It is pertinent here to mention that the whole analysis of this section can be worked out in a similar consistent fashion by working with the light-cone coordinates $y_1,y_2$. Therefore, we conclude that the two algebras are overall identical - apparent differences are solely due to the choice of coordinates. Nonetheless, they merely differ only in the relative signs of the deformation terms of (\ref{newalgebra2}) and (\ref{newalgebra}); being negative for the former and positive for the latter. Actually, these two bracket relations respectively characterize their systems and contain information on the normal mode frequencies of their respective systems. As an instance, we note that the negative sign in the modification term of (\ref{newalgebra2}) provides a footprint of the existence of different dynamical phases in a d.h.o system (over-critical, sub-critical, and critical damping phases).
%The sign in the noncommutation between boosts signifies the fact that we had, $\{ Y_i,Y_j\}=\theta\epsilon_{ij}$ and the $-$ve sign in (\ref{}) is due to the existence of the critical ratio $r=\frac{2\omega}{\lambda}$ which is responsible for the different dynamics in a D.H.O. system (viz. critical damping, oscillatory and damping, etc). The rhs of (\ref{dhosymm}) and (\ref{}) becomes same by switching to the light-cone coordinates ($Y_{\pm}=Y_1\pm Y_2$/2), without any change in the other relations \ref{}.
Therefore from the perspective of symmetries, dissipative systems obtained as a result of symplectic deformation follow that of the exotic (deformed) oscillators. This result corroborates and provides another paradigm to our previous studies in \cite{skp1}. This newly obtained unifying symmetry algebra is one of the important results of our paper. We will now identify in the following section the form of this obtained algebra as an extended version of the Newton-Hooke algebra.
\section{$AdS_3$ and extended $\overline{\mathcal{NH}}
_3$ algebra}
In this section, the extended Newton-Hooke algebra in 2+1 dimensions ($\overline{\mathcal{NH}}_3$) will be obtained as a group contraction of $AdS_3$ which is :
\begin{equation}
\{ M_{\mu \nu}, M_{\lambda \sigma} \}= (\eta_{\mu \lambda} M_{ \nu \sigma}-\eta_{\mu \sigma} M_{\nu \lambda}+\eta_{\nu \sigma}  M_{\mu \lambda}- \eta_{\nu \lambda} M_{\mu \sigma}) \nonumber
\end{equation}
\begin{equation}
\{ M_{\mu \nu}, T_{\lambda} \}= (\eta_{\mu \lambda} T_{ \nu}-\eta_{\nu \lambda}T_{\mu}) \nonumber
\end{equation}
\begin{equation}
\{ T_{\mu}, T_{\nu} \}= \Lambda M_{ \mu \nu} \nonumber
\end{equation}
where $~\mu=0,1,2~ and ~ \eta_{\mu \nu}= diag(-,+,+)$.\\
To show this, we re-write explicitly in components,
\begin{equation}
\{T_0, M_{0i}\}= T_i~;~\{M_{0i}, T_j \}= -\delta_{ij}T_0 ~;~\{T_0, T_i\}= \Lambda M_{0i} \nonumber
\end{equation}
\begin{equation}
\{ M_{0i}, M_{0j} \}= -\epsilon_{ij}M_{12} ~~;~~\{T_i, T_j\}=\Lambda \epsilon_{ij}M_{12}~;~~~ \text{where} ~i,j=1,2 \nonumber
\end{equation}
Defining the following \emph{rescaled} generators -
\begin{equation}
T_0 \longrightarrow -\xi m - \frac{1}{\xi}H \nonumber
\end{equation}
\begin{equation}
M_{0i} \longrightarrow \xi K_i~~;~~M_{12} \longrightarrow \xi^2 \sigma +J~~;~~\Lambda \longrightarrow \frac{1}{\xi^2} \Lambda \nonumber 
\end{equation}
Next, we calculate the brackets with the rescaled generators and then finally take the limit $\xi \rightarrow \infty$ to obtain - 
\begin{equation}\label{NH}
\{ H,K_i \}=-T_i ~~; ~ \{ H,T_i \}=\Lambda K_i ~;~ \{ K_i,T_j \}=m \delta_{ij} \nonumber
\end{equation}
\begin{equation}
~~~\{ K_i,K_j \}=\sigma \epsilon_{ij}~~;~~ \{ T_i,T_j \}=\Lambda \sigma \epsilon_{ij} ~~~~;~where ~i,j=1,2.
\end{equation}
This is the second centrally-extended Newton-Hooke algebra $\overline{\mathcal{NH}}_3$ which is a larger symmetry algebra. Furthermore, if one now takes the limit $\Lambda\rightarrow 0$ above, the well-known secondly extended Galilean algebra in 2+1 dimensions is obtained -
 \begin{equation}\label{Gal2}
\{ H,K_i \}=-T_i ~~; ~ \{ H,T_i \}=0 ~;~ \{ K_i,T_j \}=m \delta_{ij} \nonumber
\end{equation}
\begin{equation}
~~~\{ K_i,K_j \}=\sigma \epsilon_{ij}~~;~~ \{ T_i,T_j \}=0 ~~~~;~where ~i,j=1,2.
\end{equation}
and finally, on taking the commutative (no deformation) limit $\theta \rightarrow 0$, one arrives at the trivial firstly extended ($\emph{Gal}_3$), the Galilean algebra in 2+1 dimensions.

Now, comparing (\ref{NH}) with (\ref{boostalg1}-\ref{NCPoisson}) or (\ref{boostalg}-\ref{NCPoisson2}), we observe that the symmetry algebra for the exotic and the Bateman oscillators is similar to $\overline{\mathcal{NH}}_3$ upon identifying $\sigma=\theta m^2$ and $\Lambda=\Omega^2$. However, our reported symmetry algebra differs by the presence of an additional second term on the r.h.s. in the Lie brackets between $H$ and $K_i$ and $H$ and $T_i$. Additionally, the Lie bracket between $\{ K_i,T_j \}$ is also modified in our case, which contains information on the energy spectrum of the problem as discussed after (\ref{NCPoisson2}). In the literature, the noncommutative parameter $(\sigma)$ of the boost-boost sector (\ref{NH}) has been regarded to represent the anyonic spin in two dimensions which can take any real value \cite{jackiwnair, wilczek0, wilc}.
The radius of curvature $(R=\frac{1}{\sqrt{-\Lambda}})$ of the $AdS$ space is survived in the nonrelativistic limit by the inverse of frequency $\Omega$ of the nonrelativistic exotic (or $\tilde{\omega}$ of Bateman) oscillator and together with the deformation parameter $\theta$ is responsible for the noncommuting translations which possibly indicate to a presence of curvature in the space of motions. The Jacobi identities of the algebra have been provided in appendix B, where we demonstrate by using a general basis a mapping to the symmetry algebra.

\section{Conclusion}
We first summarize our key findings of the paper. In the beginning, we have explicitly obtained the time-dependent symmetry generators of the exotic deformed 2D oscillator through the computation of moment maps and found the associated symmetry algebra. Next, we have provided a Lagrangian that effectively describes the Bateman system (a canonical model for treating dissipative mechanics) in which the deformation parameter of position sector (in the symplectic structure) accounts for the induction of dissipation. This observation corroborates and provides fresh insights on the opinions made in the literature \cite {mairi, skp1, rbmpl} regarding a connection between dissipation and deformation through the presented Lagrangian formulation (\ref{batemanL2}). This opens up the possibility of a new paradigm in the analysis of dissipative systems in certain situations. Then we extended the analysis to compute moment map components for the modeled Bateman system to find the associated symmetries. Remarkably, the symmetry algebras are found to be almost identical in both systems. This obtained symmetry algebra has then been identified as a new kind of generalized (extended) Newton-Hooke algebra in 2+1 dimensions, $\overline{\mathcal{NH}_3}$. And importantly, on taking the zero limit of the symplectic deformation parameter $\theta$ in the obtained algebra, we recover the standard Newton-Hooke algebra in (2+1) dimensions, $\mathcal{NH}_3$. This provides a benchmark to our analysis as well because in such a limit, physically both the systems reduce to the standard two-dimensional harmonic oscillator. Thus, the present work provides a dynamical realization of the extended Newton-Hooke symmetry in terms of the dynamical models discussed. Not unexpectedly, the extended NH symmetry also consists of rotational invariance in terms of a deformed rotation generator (angular momentum) \cite{frederikapp}. However, the rotation sector of the algebra will remain unchanged and therefore has not been considered.

Most interestingly, we find that noncommuting translations (\ref{transalg1},\ref{transalg}) appear in the obtained algebra which is rather special in the usual domain of nonrelativistic spacetime. Apart from this, we obtain boost generators that do not commute with each other - a fact that is usually ascribed to anyonic spin particles. It is worthwhile to observe that the bracket relations in (\ref{newalgebra}) and (\ref{newalgebra2}) contain traces of their respective dimensionless characteristic frequencies multiplied with mass in the r.h.s. In light of all the above observations, we infer that symplectic deformations induce structural and kinematical changes in the geometry of dynamical systems viz. nonrelativistic spacetimes. One might refer to our case as generalized NH spacetimes. Previously in \cite{skp1}, we have shown that deformed oscillators have a close connection with the problem of a nonrelativistic charged particle moving in a two-dimensional plane in a background magnetic field acting normally to the plane. Our obtained algebra therefore also shares a resemblance with the reduced Galilei-Maxwell group \cite{negro06}, it being a symmetry group of the nonrelativistic charged particle moving in an electromagnetic field background and can be obtained as a specific limit of the anti-de Sitter Maxwell group \cite{gomis, gilkman}.

As mentioned before, the primary distinction between Newton-Hooke and Galilei kinematics stems from the non-commutativity between temporal and spatial translations. This leads to alterations in standard time-evolution equations where the Laplacian is time-invariant, and hence has consequences in nonrelativistic cosmology \cite{gibbons, freed, duval2017}. Therefore, this newly obtained (generalized) extended Newton-Hooke spacetimes may give rise to further and significant changes to cosmic nonrelativistic physics. It will be curious to analyze studies related to anyonic black holes in cosmological backgrounds \cite{marsot, mirza} as the symmetry algebra hints at the anyonic nature of the corresponding particles due to noncommutativity in the boost sector.

\section*{Acknowledgments}
One of the authors, S.K.P. gratefully acknowledges the support from the Indian Statistical Institute Kolkata, India for a visiting scientist fellowship, and the other author, P.N. acknowledges the support from a postdoctoral fellowship grants from Stellenbosch University, South Africa. S.K.P. extends his thanks to Subir Ghosh for several discussions and helpful suggestions during the course of this work. P.N. would like to express his appreciation to Frederik G. Scholtz for a careful reading and feedback on the manuscript.

\newpage

\section*{Appendix A: A short review on 1D H.O. symmetries}
Here, we provide a brief outline of the Galilean and the Newton-Hooke group symmetries for the one-dimensional harmonic oscillator. We recall that any transformation of a system in classical mechanics will be called a symmetry transformation iff:
 
 \begin{equation}
  \delta L= L'-L=\frac{dF}{dt}
  \end{equation}
  where F is a function of the configuration space variables.\\

Both $L$ and $L'$ give rise to the same Euler-Lagrange equations of motion. The generator $\mathcal{B}$ which gives rise to the above symmetry transformation is the conserved charge associated with the transformation and is given by the Noether prescription-
\begin{equation}
\mathcal{B}=p_i \delta q_i-F
  \end{equation}
The Hamiltonian of the 1-dim. H.O. is given by-
\begin{equation}\label{osc_normal}
H=\frac{p^2}{2m} + \frac{1}{2}m\omega^2 x^2
\end{equation}
The second-order Newton's equation of motion of the system is given by:
\begin{equation}
\ddot{x} + \omega^2 x =0
\end{equation}
which has the solution,
\begin{equation}
x= x_0\cos{\omega t}+ \frac{p_0}{m\omega}\sin{\omega t}
\end{equation}

The boost and the spatial translation generators are respectively given by the following moment maps -
\begin{equation}
K=m x \cos{\omega t}-\frac{p}{\omega} \sin{\omega t}
\end{equation}
\begin{equation}
T=m \omega x \sin{\omega t}+p\cos{\omega t}
\end{equation}
The conserved time translation generator or ``Hamiltonian" is then expressed in terms of the invariant Casimir, $H=\frac{T^2}{2m}$ following the orbit method.
Thus, the symmetry algebra of the 1D H.O. is given by:
\begin{equation}
\{K,T\}=m ~~;~~\{H,K\}=-T~~;~~\{H,T\}= 0
\end{equation}
This is a four-parameter group comprising of $K$, $T$, $H$, and the central extension $m$. All other commutators vanish. This defines the Galilean algebra in one spatial dimension ($\emph{Gal}_1$) which is the Lie algebra corresponding to the four-parameter centrally extended Galilean group, sometimes also referred to as the Bargmann algebra.
On the other hand, the Newton-Hooke $\mathcal{NH}_2$ algebra is given by the following:
\begin{equation}\label{NH2}
\{K,T\}=m ~~;~~\{H,K\}=-T~~;~~\{H,T\}=\omega^2K
\end{equation}
where the time translation generator, denoted with $H$ as before, is given by (\ref{osc_normal}). It is instructive now to write down the phase-space brackets between $x,p$, and $H$:
\begin{equation}
\{x,p\}=1~~;~~\{H,x\}=-\frac{p}{m}~~;~~\{H,p\}=m\omega^2x
\end{equation}
Replacing $K=mx$ and $p=T$ in the above reduces to Newton-Hooke algebra (\ref{NH2}), where $m$ is a constant function (central generator) to be interpreted physically as the mass. Hence, the set of Lie group generators \{K,T,H,m\} spans the Poisson sub-algebra of the Poisson algebra of well-defined functions of canonical phase-space coordinates $x$ and $p$. This provides a glimpse into the phase-space approach to symmetry principles and symmetry groups in non-relativistic physics.

\section*{Appendix B: General basis of extended planar Newton-Hooke algebra}
In this section, our objective is to define an equivalent class of the extended Newton-Hooke algebra $\overline{\mathcal{NH}_3}$ by introducing a general basis for this algebra thus providing the algebraic consistency check. Let us define,
\begin{eqnarray}
         \label{ncomdim}
             \tilde{K}_{i}&=&[a_{ij}K_{j}+b_{ij}T_{j}]~,\\
              \tilde{T}_{i}&=&[c_{ij}T_{j}+d_{ij}K_{j}]~:~~ i,j=1,2,
         \end{eqnarray}
where, $a_{ij}$ and $c_{ij}$ are dimensionless, while
$b_{ij}$ and $d_{ij}$ are dimensionful real-valued 2$\times$2 non-singular transformation matrices, respectively. Following this, the Newton-Hooke algebra is re-expressed in terms of the new generators,
\begin{equation}\label{boostalg3}
\{\tilde{K}_i,\tilde{K}_j\}= \alpha_{ij}=\sigma(a_{il}a_{jm}+\lambda b_{il} b_{jm})\epsilon_{lm}+m(a_{il}b_{jl}-b_{il}a_{jl})
\end{equation}
\begin{equation}\label{transalg3}
\{\tilde{T}_i,\tilde{T}_j\}=\beta_{ij}=\sigma(\lambda c_{ik}c_{jl}+d_{ik}d_{jl})\epsilon_{kl}+m(d_{ik}c_{jk}-c_{ik}d_{jk})
\end{equation}
\begin{equation} \label{newalgebra3}
\{\tilde{K}_i,\tilde{T}_j\}=h_{ij}=\sigma(\lambda b_{ik} c_{jl}+a_{ik}d_{jl}) \epsilon_{kl}+m(a_{ik}c_{jk}-b_{ik}d_{jk})
\end{equation}
\begin{equation}\label{NCPoisson3}
\{H,\tilde{K}_i\}=-\Sigma_{ij}\tilde{T}_{j} + \Theta_{ij} \tilde{K}_j~ ~;~\{H,\tilde{T}_i\}=\Omega_{ij}\tilde{K}_j + \Phi_{ij}\tilde{T}_j
\end{equation}
where $\Sigma_{ij}=a_{ik}C_{kj}-\lambda b_{ik}B_{kj}$, $\Theta_{ij}=\lambda b_{ik}A_{kj}-a_{ik}D_{kj}$, $\Omega_{ij}=\lambda c_{ik}A_{kj}-d_{ik}C_{kj}$, and $\Phi_{ij}=\lambda c_{ik}B_{kj}-d_{ik}C_{kj}$. The coefficients are defined as follows: $A_{ij}=-B_{ik}d_{kl}b^{-1}_{lj}$ and $C_{ij}=-D_{il}c_{lm}a^{-1}_{mj}$, and they satisfy the conditions $B^{-1}_{ij}=c_{ij}-d_{il}b^{-1}_{lk}a_{kj}$ and $D^{-1}_{ij}=b_{ij}-c_{il}a^{-1}_{lk}d_{kj}$.\\

Now, the Jacobi identities for $H, \tilde{K}_{i},\tilde{K}_{j}$, $H,\tilde{T}_{i},\tilde{T}_{j}$ and $H,\tilde{T}_{i},\tilde{K}_{j}$ are respectively given by,
\begin{equation}
    \Sigma_{jk}h_{ik}-\Theta_{jk}\alpha_{ik}-\Sigma_{ik}h_{jk}+\Theta_{ik}\alpha_{jk}=0
\end{equation}
\begin{equation}
    \Omega_{jk}h_{ij}-\Phi_{jk}\beta_{ik}-\Omega_{ik}h_{jk}+\Phi_{ik}\beta_{jk}=0
\end{equation}
\begin{equation}
\Sigma_{jk}\beta_{ik}+\Theta_{jk}h_{ki}+\Omega_{ik}\alpha_{jk}+\Phi_{ik}h_{jk}=0
\end{equation}
If we make specific choices for the structure constants as follows:
\begin{equation}
   \alpha_{ij}=\theta m^{2}\epsilon_{ij},~~\beta_{ij}=\theta m^{2}\Omega^{2}\epsilon_{ij},~~ \Sigma_{ij}=\delta_{ij},~~\Theta_{ij}=\beta \epsilon_{ij}, ~~\Omega_{ij}=\delta_{ij},~~ \Phi_{ij}=\beta \epsilon_{ij},~~h_{ij}=m(1+\frac{m^{2}\omega^{2}\theta^{2}}{2})\delta_{ij},
\end{equation}
 then the above Jacobi identities are satisfied for a suitable choice of primitive two-by-two transformation matrices. Consequently, we can assert that the algebra derived and expressed in terms of Dirac brackets (\ref{boostalg1}-\ref{NCPoisson}) falls into an equivalent class of algebra with extended Newton-Hooke symmetry written in a different basis.
\end{document}